\documentclass[conference]{IEEEtran}
\IEEEoverridecommandlockouts
\usepackage{cite}
\usepackage{amsmath,amssymb,amsfonts}
\usepackage{algorithmic}
\usepackage{graphicx}
\usepackage{textcomp}
\usepackage{xcolor}
\usepackage{todonotes}
\usepackage{graphicx}
\usepackage{stfloats}
\usepackage{placeins}
\usepackage{booktabs}
\usepackage{url}

\def\BibTeX{{\rm B\kern-.05em{\sc i\kern-.025em b}\kern-.08em
    T\kern-.1667em\lower.7ex\hbox{E}\kern-.125emX}}
\begin{document}

\title{A Defect Taxonomy for Infrastructure as Code: A Replication Study
}

\author{
    \IEEEauthorblockN{Wendell Oliveira, Filipe Paiva, Thiago Emmanuel Pereira, João Brunet}  
    \IEEEauthorblockA{Federal University of Campina Grande \\  
    \{wendell.tome.marinho.oliveira, filipe.magno.alves.paiva\}@ccc.ufcg.edu.br, \{temmanuel, joao.arthur\}@computacao.ufcg.edu.br}  
}
\maketitle

\begin{abstract} \normalfont \textbf{Background:} As Infrastructure as Code (IaC) becomes standard practice, ensuring the reliability of IaC scripts is essential. Defect taxonomies are valuable tools for this, offering a common language for issues and enabling systematic tracking. A significant prior study developed such a taxonomy, but based it exclusively on the declarative language Puppet. It remained unknown whether this taxonomy applies to programming language-based IaC (PL-IaC) tools like Pulumi, Terraform CDK, and AWS CDK. \textbf{Aim:} We replicated this foundational work to assess the generalizability of the taxonomy across a broader and more diverse landscape. \textbf{Method:} We performed qualitative analysis on 3,364 defect-related commits from 285 open-source PL-IaC repositories (PIPr dataset) to derive a PL-IaC specific defect taxonomy. We then enhanced the ACID tool, originally developed for the prior study, to automatically classify and analyze defect distributions across an expanded dataset—447 open-source repositories and 94 proprietary projects from VTEX (e-commerce) and Nubank (financial)—incorporating modern PL-IaC tools absent in the original work. \textbf{Results: } Our research confirmed the same eight defect categories identified in the original study, with idempotency and security defects appearing infrequently but persistently across projects. Configuration Data defects maintain high frequency in both open-source and proprietary codebases. Despite differences in project types, overall defect proportions remain similar. \textbf{Conclusions:} Our replication supports the generalizability of the original taxonomy, suggesting IaC development challenges surpass organizational boundaries. Configuration Data defects emerge as a persistent high-frequency problem, while idempotency and security defects remain important concerns despite lower frequency. These patterns appear consistent across open-source and proprietary projects, indicating they are fundamental to the IaC paradigm itself, transcending specific tools or project types.
\end{abstract}

\begin{IEEEkeywords}
infrastructure as code, iac defects taxonomy, open source and proprietary, idempotency and security defects, configuration data defects, replication

\end{IEEEkeywords}

\section{Introduction}
Infrastructure as Code (IaC)~\cite{iac} has established itself as an important practice in modern software development environments. By enabling the management of infrastructure through code, IaC has transformed how organizations deploy and maintain their computing resources. This approach has been widely adopted across industries, with tools such as Terraform~\cite{terraform}, Pulumi~\cite{pulumi}, AWS CDK~\cite{awscdk}, Ansible~\cite{ansible}, Chef~\cite{chef}, and Puppet~\cite{puppet} becoming standard components in DevOps toolchains~\cite{guerriero}.

Like traditional software code, IaC scripts are susceptible to defects and quality issues. These defects can propagate throughout the infrastructure, potentially leading to system failures, security vulnerabilities, and operational inefficiencies. 

Understanding the nature, patterns, and impact of these defects is important for developing effective quality assurance strategies and tooling for IaC. Some important studies have examined defects in IaC scripts to shed a light on keys aspects about their characteristics, such as, causes, fixes~\cite{georgios, defective-iac, char}, prevalence, co-occurrence, and impact~\cite{bessghaier, co-located}. In this context, three notable works stand out for providing taxonomies and categorization frameworks. The first one established a taxonomy focused on categorizing security-related defects in IaC scripts~\cite{seven}. Following this, the same research group conducted a replication study~\cite{repseven}, extending their framework to include Ansible and Chef scripts, thereby enhancing its applicability and relevance. Subsequently, in their work titled \textit{``Gang of Eight: A Defect Taxonomy for Infrastructure as Code Scripts''}, the authors expanded their research by developing a more comprehensive defect categorization framework. This framework extends beyond security concerns to encompass a broader spectrum of defect types identified through Puppet\cite{puppet} script analysis~\cite{go8}. This last work in particular is the subject of our replication.

The ``Gang of Eight'' taxonomy, while valuable, has not yet been comprehensively evaluated across emergent IaC technologies. Recently, Programming Language-based Infrastructure as Code (PL-IaC) approaches such as Terraform, Pulumi, and AWS Cloud Development Kit (AWS-CDK) have achieved significant adoption in cloud infrastructure management\cite{adoption, adoption2}. These tools utilize general-purpose programming languages including Python, Go and TypeScript to define infrastructure configurations. According Pulumi's oficial reports\cite{pulumi-report1, pulumi-report2}, between 2020 and 2023, the technology expanded its end user base from approximately 15,000 to 150,000—a tenfold increase. NPM downloads of Pulumi's core packages surged fifteen-fold during this same period. This significant adoption rate creates an opportunity to evaluate how well the framework addresses the needs of modern IaC technologies. 

Also, most existing research has focused on open-source repositories, leaving a significant gap in our understanding of how defect patterns may vary across different implementation contexts. Hence, expanding beyond open-source to include enterprise implementations would validate the existing defect categorization framework, ensuring its continued relevance to current infrastructure practices.

To bridge this gap, we replicated Rahman et al.'s study~\cite{go8} to validate both the extensibility of their defect categorization framework (through a qualitative study) and their findings on the frequency and distribution of defect types (via an empirical study). Following their original methodology, we constructed our taxonomy using the PIPr dataset~\cite{pipr}, which contains open-source PL-IaC projects. For the empirical analysis, we drew from both the PIPr dataset and proprietary projects from VTEX\footnote{https://vtex.com} and Nubank\footnote{https://www.nubank.com.br}. We selected 285 repositories from the PIPr dataset based on the original study’s inclusion criteria~\cite{munaiah}. Our proprietary sample comprised 63 projects from VTEX—a leading Latin American e-commerce platform powering over 2,000 stores in 32 countries—and 31 projects from Nubank, Brazil’s largest digital bank, serving more than 70 million customers. In total, we analyzed 94 proprietary projects. By combining PL-IaC projects from both open-source and enterprise environments, we enabled a comprehensive evaluation of the framework’s applicability across diverse technological contexts.

We conducted a manual analysis of 6,612 code commits. Of these, 3,364 commits were identified as defect-related and subjected to additional manual classification. This detailed analysis allowed us to uncover a taxonomy specific to our research context, enabling comparison with the findings from the original study. Then, we evolved ACID, the tool designed in the original study to automatically categorize defects in IaC. This enhanced version handles PL-IaC files related to Terraform, Pulumi, and AWS CDK, enabling broader applicability across other modern cloud infrastructure provisioning tools. Finally, we applied our enhanced ACID version across 541 repositories of the three datasets to address the original study's research questions regarding defect categories and their prevalence. 

Our comparative analysis confirmed that the original ``Gang of Eight'' defect taxonomy remains valid across modern PL-IaC tools and diverse environments. Configuration Data defects were consistently the most prevalent category, while Idempotency and Security defects occurred less frequently but persistently. Despite differences in project types and technologies, defect proportions were remarkably similar between the original study (15.3\%) and this replication (16.8\%). We propose a minor extension to the taxonomy by refining the Service category with two subcategories—Resource and Panic—to provide more detailed classification of service-related defects.
    
Our main contributions are:

\begin{itemize}
    \item Empirical evidence regarding the generalizability of existing defect categories across different IaC technologies and implementation environments;
    \item Comparative analysis between the original study's findings and our results across additional technologies, open-source, and proprietary projects; and
    \item An adaptation and evolution of ACID, a tool designed to automatically categorize defects in PL-IaC programs.
\end{itemize}

The remainder of this paper is structured as follows: Section~\ref{sec:background} provides background and discuss relevant related work, including the original study we are replicating. Section~\ref{sec:rqs} presents the research questions guiding our study. Section~\ref{sec:taxonomy} presents our PL-IaC defect taxonomy and the method we employed to build it, along with a comparison of our identified categories with those from the original work. Section~\ref{sec:acid-evolution} describes the evolution and validation of the ACID defect detection tool for PL-IaC programs, while Section~\ref{sec:empirical} presents the empirical results concerning the frequency and distribution of defect categories across different datasets and compares these findings to the original study. Section~\ref{sec:package} provides details on the replication package. Section~\ref{sec:threat} discusses threats to validity. Finally, Section~\ref{sec:conclusion} summarizes our conclusions and discusses future work.

\section{Background and Related Work}
\label{sec:background}

In this section, we first provide an overview of IaC, PL-IaC, and the PIPr dataset employed in our replication study. We also present a review of related work in the field, with particular emphasis on the original study we are replicating.

\subsection{Background}

\begin{figure}[!t]
  \centering
  \includegraphics[width=\linewidth]{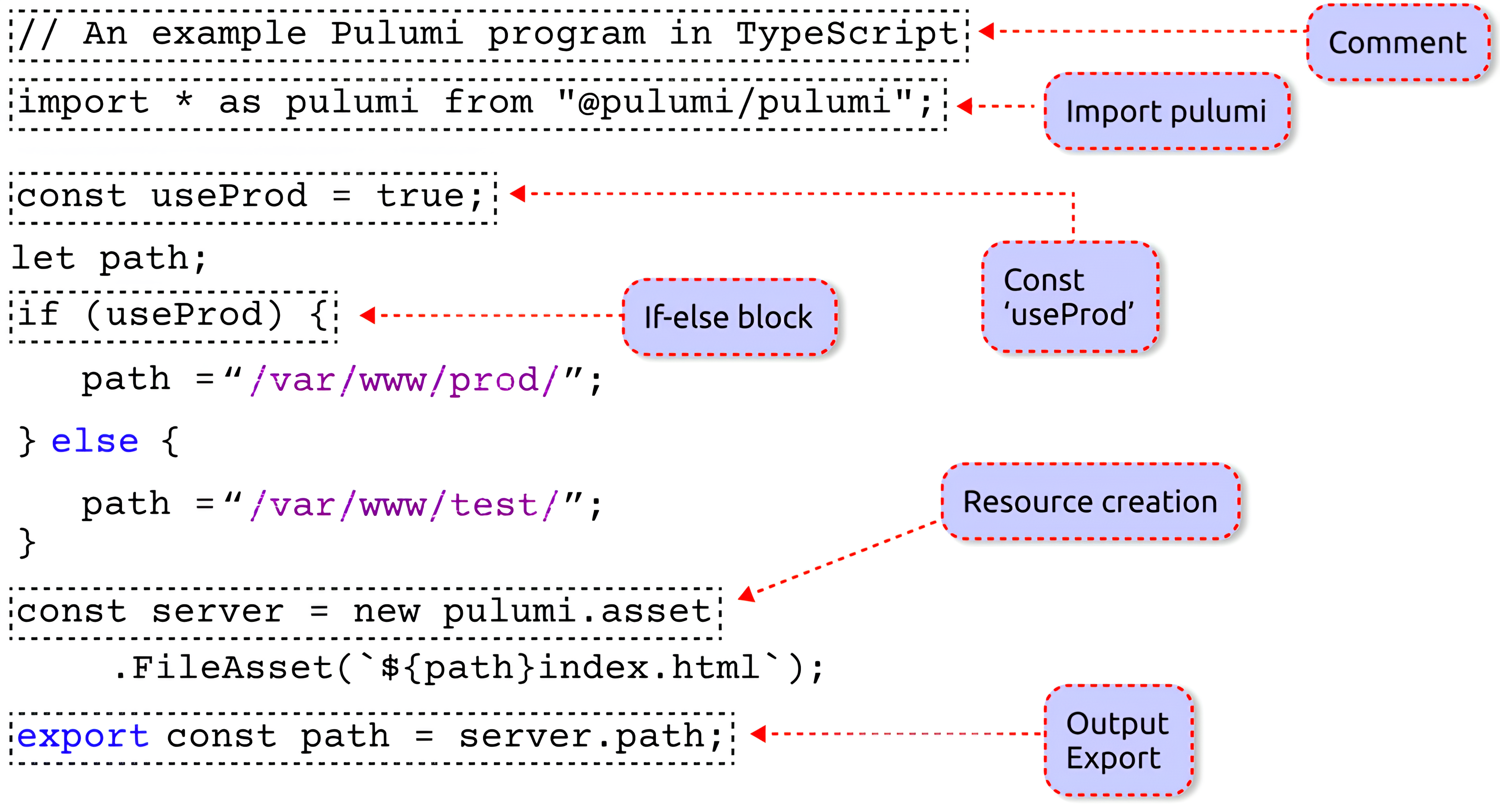}
  \caption{Example Pulumi IaC program in TypeScript}
  \label{fig:pulumi-program}
\end{figure}

\subsubsection{IaC and PL-IaC}

Infrastructure as Code (IaC) is the practice of managing and provisioning information technology (IT) infrastructure through machine-readable definition files~\cite{iacbook}. Generally, IaC scripts are written using declarative domain-specific languages (DSLs) or configuration languages such as JSON and YAML. In contrast, Programming Language-based Infrastructure as Code (PL-IaC) makes use of general-purpose programming languages (e.g., TypeScript, Python), allowing developers to utilize full programming language capabilities for infrastructure definition through IaC programs instead of IaC scripts~\cite{pipr}. We provide a sample Pulumi IaC program in Typescript with annotations in Figure~\ref{fig:pulumi-program}. Note that the code incorporates typical programming constructs such as conditionals (if-else) and variables, enabling developers to describe how to achieve the desired infrastructure state through familiar coding paradigms rather than declarative specifications.

This paradigm shift has driven measurable improvements: a report from Deepwatch company stated that they migrated a complex backend infrastructure in less than two weeks using AWS CDK, achieving a 60\% efficiency gain through CI/CD integration and reusable patterns~\cite{awscdkimprove}. Similarly, Panther Labs~\cite{pantherlabs2025} reduced its codebase by more than 50\% and accelerated deployments 10x by adopting Pulumi~\cite{pantherimprove}. Notable examples of PL-IaC solutions include AWS CDK~\cite{awscdk}, CDKTF~\cite{cdktf}, and Pulumi~\cite{pulumi}.

\subsubsection{PIPr Dataset}

For the development of the taxonomy in our replication, we utilize The PIPr Dataset of Public Infrastructure as Code Programs, introduced by Sokolowski et al.~\cite{pipr}. This dataset represents the first systematic collection specifically dedicated to PL-IaC programs and was constructed through the mining of public repositories on GitHub, and as of August 2022, it encompasses metadata for 37,712 IaC programs distributed across 21,445 repositories. Furthermore, PIPr contains the source code of 15,504 IaC programs for which the licensing conditions permit redistribution.

The selection of the PIPr dataset for this study is based on several factors. Primarily, it maintains a focus on the PL-IaC paradigm, directly aligning with the goal of our replication. PIPr offers a comprehensive and recent snapshot of public PL-IaC programs, enabling analysis of current defect patterns in the field. Additionally, as the first systematic dataset in the domain of PL-IaC, PIPr establishes a foundational resource for empirical research in this area. Despite its relatively recent development in 2024, PIPr has already been utilized by multiple research studies \cite{iac-testing}\cite{qiao}~\cite{limp}.

\subsection{Related Work}

Over recent years, researchers have investigated defects and faults in IaC scripts. Drosos et al.~\cite{georgios}, for example, examined 360 bugs across three IaC ecosystems to understand their characteristics, causes, and fixes while providing recommendations for improved reliability tools. Bessghaier et al.~\cite{bessghaier} investigated the prevalence, co-occurrence, and impact of Infrastructure-as-Code smells, finding that 74\% of files contain quality issues which lead to nearly 4 times more modifications. Another work, conducted by Rahman and Williams~\cite{defective-iac}, applied text mining techniques on IaC scripts from Mozilla, Openstack, and Wikimedia Commons to identify characteristics of defective code and build defect prediction models to help practitioners prioritize validation and verification efforts.

Building upon this landscape of IaC research, we focus on three studies that have established a foundation for understanding defects in this domain and directly influence our approach and methodology. The progression of these characterizations follows a logical timeline; therefore, we present these particular related works in chronological order to show how research in this field led to our current replication study.

Rahman et al.~\cite{seven} initially focused on categorizing security-specific defects in IaC scripts. Their seminal work identified patterns and categories of security smells analyzing 15,232 IaC scripts from 293 open-source repositories. The study identifies seven critical security smells —such as hard-coded secrets and weak cryptography—revealing their prevalence (21,201 occurrences in 15,232 scripts), long persistence (up to 98 months), and practitioner acknowledgment (64\% agreement on fixes). 

The same research group further demonstrated the applicability of their approach by replicating their security defect study across multiple IaC languages beyond those examined in their original work~\cite{repseven}. This cross-language validation provided evidence that certain defect patterns transcend specific IaC implementation technologies, while others may be language-specific. 

Building upon this previous work, Rahman and colleagues expanded their focus to develop a more comprehensive taxonomy of IaC defects beyond security concerns~\cite{go8}. This broader framework, presented in their subsequent study, categorizes defects across multiple dimensions including configuration errors, dependency issues, idempotency problems, and resource management flaws. \textbf{This expanded taxonomy forms the basis for our current replication study.} The authors applied descriptive coding to 1,448 defect-related commits from OpenStack repositories to identify the categories. These categories were validated with 66 practitioners to assess agreement with the taxonomy. Then, they built a rule-based tool, named ACID, using dependency parsing and pattern matching to classify defects in 80,415 commits from 291 OSS repositories to quantify defect frequency and evolution over time.

Despite the relevance of establishing a comprehensive, empirically validated taxonomy of defects in IaC, the authors opened up questions to be answered by future works. In particular, they explicitly pointed out that, because they built the taxonomy upon declarative Puppet scripts, it may not generalize for PL-IaC programs that use an imperative form of language. In addition, they also acknowledged their research used only open-source software scripts, not proprietary sources, creating potential gaps in their findings.

Our replication study builds upon this established research trajectory by examining how this taxonomy applies across diverse technological environments, including both open-source and proprietary projects, thereby investigating the generalizability of the original findings.

\section{Research Questions}
\label{sec:rqs}

The original study addressed two research questions related to the categories and frequency of IaC defects: i) RQ1:\textit{ What categories of defects occur in IaC scripts?}; and ii) \textit{RQ2: How frequently do these defect categories appear in IaC scripts?} In our study, we aim to answer these same questions in the context of PL-IaC. Accordingly, we adapted them as follows:

\begin{itemize}
    \item \textbf{RQ1:} What categories of defects appear in PL-IaC programs?
    \item \textbf{RQ2:} How frequently do the identified defect categories appear in PL-IaC programs?
\end{itemize}

Besides addressing the original RQs, we also reasoned on the differences and similarities of our findings. For this reason, we formulated the additional following RQs:

\begin{itemize}
    \item \textbf{RQ1.1:} To what extent are the defect categories identified in our context (different IaC languages and projects) similar to or different from those documented in the original study?
    \item \textbf{RQ2.1:} How do the frequency and distribution of PL-IaC defect types compare with those observed in the original study?
\end{itemize}

We omitted addressing the original study's question about practitioner perceptions of IaC defect categories for two key reasons: First, IaC importance and development practices are now well-established across industry and academia, making additional developer surveys unlikely to yield significantly new insights beyond existing documentation. Second, during our validation and tool development process, 22 participants indirectly validated our categorization framework by not raising concerns or suggesting alternatives throughout the extensive review, providing sufficient confidence in our approach without requiring a dedicated research question on practitioner perception.

To answer these research questions, we followed the methodology of the original study as closely as possible, keeping the same procedures and analysis methods. Throughout the next sections, we describe our approach with reference to the original study, clearly pointing out the few differences where we had to make adjustments. These changes were made only when necessary for our specific context, while maintaining the core replication integrity. By following the original methodology closely, we were able to properly compare our findings with the original study and help verify its conclusions.

\section{Defect Taxonomy for PL-IaC}
\label{sec:taxonomy}

In this section we present the taxonomy we built from our analysis (RQ1) along with the comparison of the categories we found against the ones present in the original study (RQ1.1). For each research question, we first detail the methodology used to conduct the analysis.

\subsection{What categories of defects appear in PL-IaC programs?}

\begin{figure*}[!t]
  \centering
  \includegraphics[width=\textwidth]{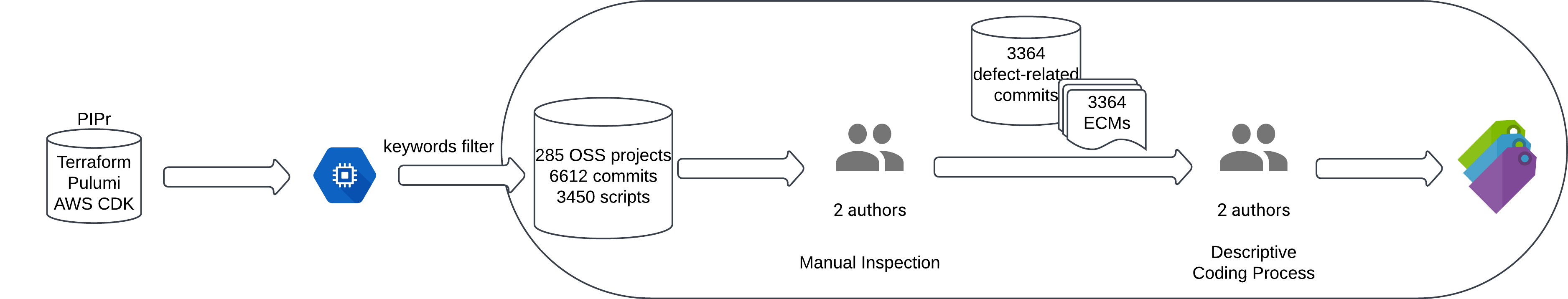}
  \caption{Methodology to Develop Defect Taxonomy}
  \label{fig:categorization}
\end{figure*}

\subsubsection{Method} 
Figure~\ref{fig:categorization} illustrates the qualitative process employed in OSS projects from the PIPs dataset for categorizing defects. This process is identical to the one utilized in the original study that we are replicating, ensuring consistency and comparability of results.

First, we used the same keywords as the original study to filter 6,612 commits from a sample of 285 OSS projects hosted in the PIPr dataset. The repository selection process is identical to the original study and based on guidelines provided by Munaiah et al.~\cite{munaiah} for curation of OSS repositories before conducting empirical analysis. First, repositories must contain a minimum threshold of 11\% PL-IaC programs relative to total files. Second, repositories must be original work rather than clones of existing repositories. Third, repositories must demonstrate consistent development activity with at least two commits per month. Fourth, repositories must have substantial collaborative input from at least 10 different contributors.

The code was identified by specific file indicators: Pulumi code resides in directories with ``Pulumi.yaml/yml'' files, AWS CDK in directories with ``cdk.json'' files, and Terraform either has ``.tf'' extensions or is in directories with ``cdktf.json'' files. All implementations' artifacts must use officially supported languages for their respective platforms.

Then, the first and second authors manually inspected the 6612 commits to select the defect-related ones. This process resulted in 3,364 defect-related commits and their corresponding Enhanced Commit Messages (ECMs), which combine the original commit message with any associated bug report descriptions.

Table~\ref{tab:comparison} compares the dataset numbers of our replication study with the original research. While filtering fewer total commits (6,612 versus 7,808), our study significantly expanded both project diversity (285 OSS projects compared to 61) and code coverage (3,140 versus 1,722). This broader analytical scope yielded 3,364 defect-related commits (ECMs), representing a 17\% increase over the original study.

\begin{table}[ht]
\centering
\caption{Comparison between Original Study and Our Study}
\label{tab:comparison}
\begin{tabular}{lrrrr}
\hline
\textbf{Study} & \textbf{OSS Proj.} & \textbf{Commits} & \textbf{Scripts/PL-IaC} & \textbf{ECMs} \\
\hline
Original Study & 61 & 7,808 & 1,386 & 1,448 \\
Our Study & 285 & 6,612 & 3,450 & 3,364 \\
\hline
\end{tabular}
\end{table}

The last step was to categorize the defect-related PL-IaC programs. The descriptive coding process follows a structured three-stage methodology. In the first stage, the first and second authors analyze the ECM of each commit that modifies the code. From these messages, they extract text segments that describe either the reason for the defect or its symptoms, capturing this information as raw text. The second stage involves generating initial categories from the extracted raw text. These preliminary classifications represent the researchers' first attempt to group similar defects based on their characteristics. The third stage consists of refining these initial categories by combining those that represent the same underlying pattern. This consolidation process results in a final set of defect categories that accurately represent the patterns observed across the codebase.

To ensure methodological rigor, the two researchers independently conducted this analysis. The Cohen’s Kappa~\cite{kappa} for the categorization was 0.78 (original study was 0.8), indicating substantial inter-rater agreement~\cite{landis} on the identified defect categories. One researcher identified two additional categories not recognized by the other: concurrency errors and type errors. Similarly, the second researcher identified a unique category—cache errors—that was not captured by the first researcher's classification. To resolve these discrepancies, the researchers discussed and determined that the type errors category substantively overlapped with syntax errors, a category both researchers had previously identified. Consequently, these were consolidated under the existing syntax errors classification. Also, the concurrency errors category was mapped to configuration data errors, another category both researchers had independently identified. The cache errors category was thoroughly evaluated and subsequently incorporated into the final taxonomy after determining its distinct characteristics warranted separate classification.

\subsubsection{Answer to RQ1} 

Our analysis revealed 8 categories. They are: \textit{Cloud, Configuration, Dependency, Documentation, Idempotency, Logic, Security, and Syntax.} 

\textit{Cloud} defects involve problems provisioning or managing services like databases or web servers. \textit{Configuration} defects involve incorrect settings or values, while \textit{Dependency} errors arise from missing or improperly specified external requirements like packages or files. \textit{Documentation} defects refer to inaccurate information in comments or READMEs. A category specific to this context is \textit{Idempotency}, where code causes unintended changes when run multiple times instead of ensuring a consistent state. \textit{Conditional} defects relate to flawed logic or branching conditions, while \textit{Security} defects introduce vulnerabilities such as exposing secrets. \textit{Syntax} defects are basic grammatical errors in the code.

\begin{figure}[!t]
  \centering
  \includegraphics[scale=0.26]{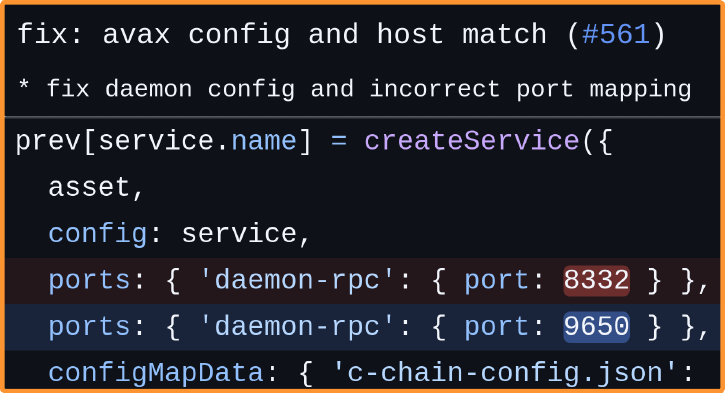}
  \caption{Incorrect port mapping.}
  \label{fig:example-configuration-data-network}
\end{figure}

Syntax represents the most prevalent category in our manual classification, with 933 defects identified within it, followed by Configuration and Dependency, with 895 and 613 defects, respectively. Cloud and Documentation categories both contain 407 defects. Security comprised 158 defect, while we mapped 50 and 7 defects to Logic and Idempotency, respectively. Figure~\ref{fig:example-configuration-data-network} provides an example of a Configuration defect reported and fixed by a practitioner in the shapeshift/unchained repository\footnote{https://github.com/shapeshift/unchained/commit/\\725c867a181ab85241f37fc3e989a1aedc1ccd83}. The figure shows a code diff where the port number for the daemon-rpc service is changed from '8332' to '9650'. This type of modification alters the service's network configuration, impacting how components communicate and how external systems connect.

\textbf{Cloud subcategories.} During the classification process, we observed a recurring defect pattern: in several cases, the provisioned infrastructure was well-structured and seemingly correct, yet failed post-deployment validations due to subtle, environment-specific conditions. For example, an AWS health check for an EC2 instance failed due to an edge case involving CPU limit configurations, despite the instance being correctly provisioned according to the specification~\footnote{\url{https://github.com/adityamillind98/checkov/commit/c2092e760ff07639313c191e23c03459227dea12}}. 

Further investigation revealed similar reports in community discussions, such as a Stack Overflow post\footnote{https://stackoverflow.com/questions/78786994/aws-auto-scaling-group-health-check-grace-period-value-reset?rq=1} where an Auto Scaling Group did not pass AWS health checks due to an insufficient grace period configuration. Although the infrastructure components were technically correct, the default timing behavior led to unintended instance reboots or failures during service checks.

Based on these findings, we refined the \textbf{Cloud} category by introducing two subcategories:
\begin{itemize}
    \item \textit{Resource:} Includes issues related to misconfigured or misplaced infrastructure resources;
    \item \textit{Panic:} Captures edge-case defects that trigger unexpected runtime behavior (e.g., reboots, failed health checks, or kernel crashes) despite configurations being syntactically and structurally correct.
\end{itemize}

\textbf{Configuration subcategories.} Configuration represents a major and broad concern in IaC context. For this reason, the following specific subcategories for different configuration defects were identified:
\begin{itemize}
    \item \textit{Cache:} defects related to caching mechanisms;
    \item \textit{Credential:} defects related to user credentials; 
    \item \textit{FileSystem:} issues including incorrect paths and permissions, and directory structure problems;
    \item \textit{Network:} defects in network configurations, protocols, connectivity, and API endpoints.
    \item \textit{Storage:} problems with data persistence, including database configuration errors and storage resource allocation.   
\end{itemize}


\subsection{To what extent are the defect categories identified in our context (different IaC languages and projects) similar to or different from those documented in the original study?}

\subsubsection{Method} 
We implemented a straightforward analytical process, directly comparing the defect categories identified in our study against those documented in the original research, mapping the names when necessary. This comparative analysis enabled us to  evaluate the consistency and validity of the established defect taxonomy across different contexts.

\subsubsection{Answer to RQ1.1} 

In a nutshell, after mapping the names, we identified \textbf{the same eight defect categories} described in the original study --- Conditional, Configuration Data, Dependency, Documentation, Idempotency, Security, Service, and Syntax --- across our broader PL-IaC dataset. This validates the generalizability Rahman et. al taxonomy to modern IaC technologies such as Terraform, Pulumi, and AWS CDK. 

Three of our categories, despite having different names, are semantically equivalent to those in the original study. We mapped them without compromising our taxonomy. Our Logic category corresponds to Conditional in the original study, while our Configuration category aligns with Configuration Data. We classified cloud service-related defects as Cloud, which maps to the Service category in the original study. From now on, we will use the terms chosen by the original work, i. e., Conditional, Configuration Data, and Service.

The only significant difference is the two subcategories we found for Service (Cloud before mapping). We believe Resource and Panic subcategories can be applied to give more details to Service defects. However, in our opinion, this refinement does not imply a departure from the original taxonomy, but rather a minor extension proposal.

\section{ACID Evolution and Validation}
\label{sec:acid-evolution}

Before conducting an empirical analysis on the presence of the categorized defects in a broader dataset, we first needed to enhance and validate the tooling used for automatic detection and categorization of defects. Hence, we have adapted the original study tool\footnote{Our enhanced version of the ACID tool is available at https://figshare.com/s/22f9887a3c3b318a67d5}, named ACID, which automatically detects and classifies defects based on the established categories.

We report this process in this section. First, we describe the process of evolving the original defect detection tool to accommodate refined detection rules based on our preliminary analysis. Second, we detail the validation methodology, which mirrors the approach of the original study by employing 22 independent raters to establish an oracle dataset for evaluating the enhanced tool's performance. Third, we report the results of our refined ACID implementation validation.

\begin{figure}[!t]
  \centering
  \includegraphics[width=\linewidth]{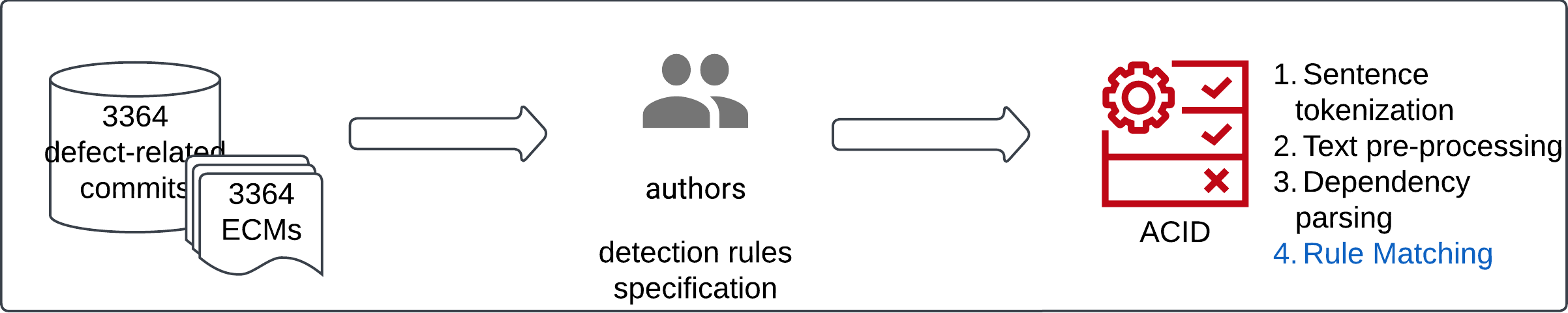}
  \caption{ACID Evolution}
  \label{fig:acid-evolution}
\end{figure}

\subsection{Evolving ACID rule set} 

Figure \ref{fig:acid-evolution} illustrates the process we carried out to evolve ACID rules to automatically classify defects. We preserved most of the original workflow while specifically targeting the rule matching component. In this enhancement process, we identify defect categories in code changes through pattern analysis. The first and second authors manually reviewed all 3,364 defect-related ECMs identified during the taxonomy construction to predefine rules that capture patterns such as ``fix'' in combination with related terms (e.g., ``warning'', ``security''). The process includes: Extracting defect-related actions and their dependencies from commit messages; Analyzing code diffs when commit messages lack clarity; Applying rule-based detection using string pattern matching; and Supporting multi-category classification when multiple defect patterns appear. The detailed description of these steps is presented in the original paper.

ACID then employs a four-step process to identify defect categories: sentence tokenization, text pre-processing, dependency parsing, and rule matching. Our modifications focused on the Rule Matching component, which identifies defect patterns using the predefined rules. We updated this component with new detection rules while preserving the structure of the previous steps in the ACID workflow. 

Table~\ref{tab:patterns} shows the patterns we found and maps to the functions of our rule set. For example, the hasDefect() function activates when specific keywords from the table appear in a commit message sentence, flagging it as a potential defect.

Table~\ref{tab:rules} presents our set of rules derived from the ECMs. We generated a rule set similar to the original study but implemented some modifications to increase accuracy. We removed 4 rules and improved 3 others to reduce false positives. Removed Rules are:

\begin{itemize}
    \item Documentation: Removed ``spec'' due to conflicts with ``specification'' that had no documentation relationship and removed ``header'' due to conflicts with ``Security headers'';
    \item Configuration Data Network: removed ``ip'' due to conflicts with words containing ``ip'' within other contexts and removed ``address'' due to conflicts with ``addressing'' terms not indicating network issues.
\end{itemize}

Rules changed are:

\begin{itemize}
    \item Security: Changed ``secur'' to ``security'' (avoided false positives from phrases like ``secure connection'' unrelated to security defects);
    \item Dependency: Changed ``compat'' to ``compatibil'' (avoided false positives from phrases like ``compatible format'' unrelated to dependency issues);
    \item  Conditional: Changed ``condit'' to ``condition'' for more precise matching.
\end{itemize}

\begin{table}[ht]
\centering
\caption{String Patterns Used for Functions in Rules}
\label{tab:patterns}
\begin{tabular}{@{}lp{6cm}@{}}
\toprule
\textbf{Function} & \textbf{String Pattern}                                                                                                                                 \\ \midrule
hasDefect()       & 'error', 'bug', 'fix', 'issu', 'mistake', 'incorrect', 'fault', 'defect', 'flaw','solve'                                                               \\
hasCond()         & 'logic', 'condition', 'boolean'                                                                                                                         \\
hasStorConf()     & 'sql', 'db', 'databas', 'disk'                                                                                                                          \\
hasFileConf()     & 'file', 'permiss'                                                                                                                                       \\
hasNetConf()      & 'network', 'port', 'tcp', 'dhcp', 'ssh', 'gateway', 'connect', 'rout'                                                                                   \\
hasCredConf()     & 'user', 'usernam', 'password', 'polic', 'credential', 'iam', 'role', 'token'                                                                            \\
hasCachConf()     & 'cach', 'memory', 'buffer', 'evict', 'ttl'                                                                                                              \\
hasDepe()         & 'requir', 'depend', 'relation', 'order', 'sync', 'compatibil', 'ensure', 'inherit', 'version', 'deprecat', 'packag', 'path', 'modul', 'upgrad', 'updat' \\
hasDoc()          & 'doc', 'comment', 'licens', 'copyright', 'notic', 'readm', 'descript'                                                                                   \\
hasIdem()         & 'idempot', 'determin'                                                                                                                                   \\
hasSecu()         & 'vulner', 'ssl', 'secr', 'authent', 'password', 'security', 'cve', 'cert', 'firewall', 'encrypt', 'protect', 'access'                                    \\
hasServResour()   & 'servic', 'server', 'location', 'resourc', 'provi', 'cluster'                                                                                           \\
hasServPanic()    & 'check', 'deploy', 'reboot', 'build', 'mount', 'kernel', 'extran', 'bypass'                                                                             \\ \bottomrule
\end{tabular}
\end{table}

Furthermore, in Table~\ref{tab:rules}, we retain the original detection functions --- \textit{changedInclude(x.diff)}, \textit{changedComment(x.diff)}, \textit{changedService(x.diff)}, and \textit{dataChanged(x.diff)} --- which identify changes in include statements, comments, service resources, and configuration data, respectively. Based on our analysis, we introduce three new functions: \textit{dataNetChanged(x.diff)}, \textit{dataCredChanged(x.diff)}, and \textit{changedSecu(x.diff)}, which detect changes related to networking configurations, credentials configurations, and security statements. While retaining the original \textit{changedComment(x.diff)} function, we refined its implementation to better align with comment syntax patterns in \textit{PL-IaC} based on each documentation. 


\begin{table}[htbp]
\centering
\caption{Rules to Detect Defect Categories}
\label{tab:rules}
\begin{tabular}{@{}lp{6cm}@{}}
\hline
\textbf{Category} & \textbf{Rule} \\ \hline
Conditional & hasDefect(x.sen) $\wedge$ hasCond(x.sen.dep) \\ \hline
Configuration Data & hasDefect(x.sen) $\wedge$ ((hasStorConf(x.sen.dep) $\vee$ hasFileConf(x.sen.dep) $\vee$ (hasNetConf(x.sen.dep) $\vee$ dataNetChanged(x.diff)) $\vee$ (hasCredConf(x.sen.dep) $\vee$ dataCredChanged(x.diff)) $\vee$ hasCachConf(x.sen.dep)) $\vee$ dataChanged(x.diff)) \\ \hline
Dependency & hasDefect(x.sen) $\wedge$ (hasDep(x.sen.dep) $\vee$ changedInclude(x.diff)) \\ \hline
Documentation & hasDefect(x.sen) $\wedge$ (hasDoc(x.sen.dep) $\vee$ changedComments(x.diff)) \\ \hline
Idempotency & hasDefect(x.sen) $\wedge$ hasIdem(x.sen.dep) \\ \hline
Security & hasDefect(x.sen) $\wedge$ (hasSecu(x.sen.dep) $\vee$ changedSecu(x.diff)) \\ \hline
Service & hasDefect(x.sen) $\wedge$ ((hasServResourc(x.sen.dep) $\vee$ changedService(x.diff)) $\vee$ hasServPanic(x.sen.dep)) \\ \hline
Syntax & hasDefect(x.sen) $\wedge$ hasSyntax(x.sen.dep) \\ \hline
\end{tabular}
\end{table}

\subsection{Building an Oracle to validate ACID new version} 
\begin{figure}[!t]
  \centering
  \includegraphics[width=\linewidth]{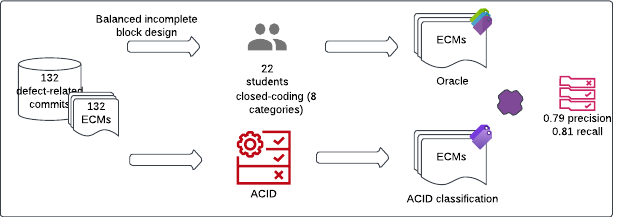}
  \caption{ACID Validation}
  \label{fig:acid-validation}
\end{figure}

We rigorously followed the methodological steps established by Rahman et al. in their original study. Figure~\ref{fig:acid-validation} illustrates these steps. We established an oracle dataset through 22 independent raters who were not authors of this paper, maintaining the exact number of evaluators as in the original research. These raters performed closed coding~\cite{coding} to map the defects to our identified categories using a structured review process. We recruited the 22 volunteers from graduate and undergraduate level DevOps practitioners and utilized a balanced incomplete block design to distribute commits among raters, ensuring each commit was reviewed by at least two individuals. Upon completion, disagreements were resolved through arbitration by the first author to finalize the oracle dataset.

\subsection{Validation results} We then evaluated our enhanced version of ACID tool by computing precision and recall metrics against this oracle dataset. The evaluation demonstrated an average precision of 0.79 and recall of 0.81 across all defect categories. It showed lower precision and recall values compared to the original study's metrics (0.84 and 0.9); these results still demonstrate robust detection capabilities adequate for reliable defect classification.

\section{Empirical Study}
\label{sec:empirical}

Having established the PL-IaC defect taxonomy (Section IV) and validated the enhanced ACID detection tool (Section V), we now transition to the quantitative analysis of defect prevalence. This section presents the empirical findings derived from applying the ACID tool across our comprehensive datasets (PIPr, VTEX, Nubank). We specifically address RQ2, examining the frequency and distribution of the identified defect categories in PL-IaC programs, and RQ2.1, comparing these findings against the original study.

\subsection{How frequently do the identified defect categories appear in PL-IaC programs?}

\subsubsection{Method} 
To address RQ2, we use the exact same process and metrics of the original study. In a nutshell, we measured defect prevalence using three metrics: ``defect proportion'' (percentage of commits containing a specific defect type), ``script proportion''~\footnote{We kept the term script to be consistent with the original work. However, in our case, it means PL-IaC programs.} (percentage of code artifacts modified in commits with at least one defect category), and ``defect/year'' (temporal trend analysis).

\textbf{Dataset.} The PIPR dataset, which we described earlier, served as our primary research corpus. To extend our analysis to corporate repositories, we incorporated 94 repositories from two large companies: VTEX and Nubank. For the corporate projects, we excluded one of the four criteria described earlier: having at least 10 different contributors. While this criterion is important for open source projects to ensure substantial collaborative work, VTEX and Nubank have dedicated teams for DevOps operations that typically consist of fewer than 10 members. Including this requirement would have excluded important repositories from our sample, as pointed out by the VTEX developer who had collaborated with us in this study.

Table~\ref{tab:dataset} presents the key attributes of these three datasets, providing a comparison of their size, scope, and characteristics. The last column compares our numbers with the original study. A significant aspect of our dataset is the diversity of PL-IaC technologies. The 38,897 PL-IaC programs from our repositories include Terraform, Pulumi, and AWS CDK, with the general-purpose programming languages used to define and manage infrastructure configurations. The most common languages in our dataset are TypeScript (7,461), C\# (4,822), and Python (3,170). Other less frequent languages include Go, Javascript, Java, Visual Basic, and F\#. This diverse dataset enabled us to evaluate defect pattern consistency across different organizational contexts, development environments, and technologies, improving the generalizability of our findings.

\begin{table}[htbp]
\centering
\caption{Attributes of the Three Datasets}
\label{tab:dataset}
\begin{tabular}{lrrr|r}
\hline
\textbf{Att} & \textbf{PIPr} & \textbf{VTEX} & \textbf{NUB.} & \textbf{Tot. (Ours / Go8)} \\ \hline
Repos       & 447      & 63       & 31       & 541 / 291      \\
Commits     & 503,585  & 15,272   & 49,834   & 570,491 / 564,947  \\
Scripts/Programs & 31,855   & 1,423    & 5,619    & 38,897 / 17,609   \\
IaC Commits & 26,399   & 2,068    & 7,615    & 36,082 / 66,423   \\ \hline
\end{tabular}
\end{table}

\textbf{Sanity Check.} Prior to our empirical analysis, we manually validated our ACID version using 2,000 randomly selected extended commit messages (1,000 from PIPr and 1,000 from VTEX). This matches the sample size used in the original work. For privacy reasons, we could not select samples from Nubank's code for manual analysis. Table~\ref{tab:sanity-checking-acid} details precision and recall for each category. On average, our ACID version achieved 0.88 and 0.87, precision and recall, respectively. These numbers confirm ACID's strong detection capabilities while acknowledging some false negatives and false positives. The main area of disagreement (15 occurrences) involves items the tool marked as No Defect, while we categorize primarily as Configuration Data, Dependency, or Syntax. Another important difference (12 occurrences) is the tool flagging Documentation issues that we considered No Defect. 

\begin{table}[htpb]
\centering
\caption{Sanity Checking ACID's Accurancy}
\label{tab:sanity-checking-acid}
\begin{tabular}{lrrr}
\hline
\textbf{Category}  & \textbf{Occur.} & \textbf{Precision} & \textbf{Recall} \\ \hline
Conditional        & 86              & 0.71               & 1.00            \\
Configuration Data & 204             & 0.93               & 0.86            \\
Dependency         & 119             & 0.92               & 0.76            \\
Documentation      & 103             & 0.62               & 0.96            \\
Idempotency        & 1               & 1.00               & 1.00            \\
Security           & 8               & 1.00               & 0.80            \\
Service            & 70              & 0.94               & 0.81            \\
Syntax             & 47              & 0.87               & 0.67            \\ \hline
No Defect          & 1,362           & 0.97               & 0.98            \\ \hline
Average            &                 & 0.88               & 0.87            \\ \hline
\end{tabular}
\end{table}

\subsubsection{Answering RQ2}

All metrics reported in our datasets were computed using the same Python and R scripts provided by the original study, available via a Docker image~\footnote{https://hub.docker.com/r/akondrahman/acid-puppet}. This ensured the reproducibility of the original analysis, including figures and metric calculations.

We report the defect proportions and script proportion for the eight categories in Table~\ref{tab:defect-script-proportion}. Configuration data is the most frequent category in all datasets, affecting more than \textbf{30\%} of Nubank's PL-IaC programs. Overall, \textbf{15.8\% to 32.8\%} of the PL-IaC programs are modified in commits related to this category---the highest among all types of defects. As shown in the 'Total' row, \textbf{18.1\% (PIPr)}, \textbf{16.1\% (VTEX)}, and \textbf{16.4\% (Nubank)} of all commits include at least one defect. In terms of scripts, \textbf{26.4\%}, \textbf{20.3\%}, and \textbf{38.7\%} respectively contain at least one defect. Consistent with the defect proportions, Configuration Data also dominates in script proportions across all datasets.

Among the subcategories of Configuration Data, network is the most frequent, accounting for 61.91\%–65.84\% of the identified defects. It is followed by credential-related defects (30.21\%–42.77\%), file system (3.61\%–14.07\%), storage (2.41\%–4.39\%), and cache (0.19\%–1.81\%). For the Service category, the resource subcategory is more frequent (77.78\%–85.75\%), while panic defects appear less often (18.23\%–24.07\%).

\begin{table}[htbp]
\caption{Defect and Script Proportion for Defect Categories}
\label{tab:defect-script-proportion}
\centering
\begin{tabular}{lrrr|rrr}
\hline
               & \multicolumn{3}{c|}{\textbf{Defect Prop. (\%)}} & \multicolumn{3}{c}{\textbf{Script Prop. (\%)}}  \\ \hline
\textbf{Categ} & \textbf{PIPr} & \textbf{VTEX} & \textbf{NUB.} & \textbf{PIPr} & \textbf{VTEX} & \textbf{NUB.} \\ \hline
Cond.          & 6.0           & 4.3           & 2.2             & 9.3           & 4.8           & 6.1             \\ \hline
Conf.Data      & 10.1          & 10.5          & 12.1            & 15.8          & 17.2          & 32.8            \\ \hline
Depe.          & 9.8           & 7.4           & 3.2             & 14.0          & 8.1           & 9.5            \\ \hline
Docu.          & 6.5           & 5.3           & 4.5             & 10.8          & 6.5           & 10.4            \\ \hline
Idem.          & 0.1           & 0.0           & 0.0             & 0.2           & 0.0           & 0.0             \\ \hline
Secu.          & 0.6           & 0.4           & 0.4             & 1.3           & 0.8           & 1.9             \\ \hline
Serv.          & 6.3           & 2.6           & 2.5             & 10.5          & 5.4           & 8.6            \\ \hline
Synt.          & 3.3           & 2.3           & 2.2             & 8.4           & 4.5           & 7.9             \\ \hline
\textbf{Total} & 18.1          & 16.1          & 16.4            & 26.4          & 20.3          & 38.7            \\ \hline
\end{tabular}
\end{table}


Defect categories are not mutually exclusive; ECMs can satisfy multiple classification rules, as also reported in the original study. ACID supports this by reporting multilabeled ECMs. We observe that 10.83\% (PIPr), 8.27\% (VTEX), and 15.60\% (Nubank) of ECMs are associated with two defect categories. Multicategory cases persist at higher levels: three categories in 7.84\%, 5.40\%, and 5.80\% of ECMs, and four in 4.70\%, 2.38\%, and 2.51\%, respectively.

\begin{figure*}[!t]
  \caption{Evolution of defect proportion for eight categories; PIPr, Nubank, VTEX. For each dataset, ``Total'' presents the proportion of commits, which includes at least one category of defect.
    The colored lines in the graph represent curve fits applied to the data points. The gray shaded area indicates the 95\% confidence interval, showing where the true trend would likely lie if the data were collected repeatedly.
  }
  \centering
  \label{fig:evolution-defect-year}
  \textbf{PIPr}
  \includegraphics[width=\linewidth]{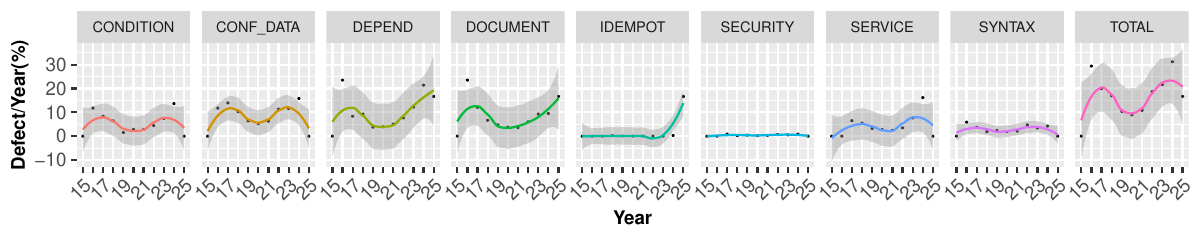}
  \label{fig:pipr-defect-year}

  \textbf{VTEX}
  \includegraphics[width=\linewidth]{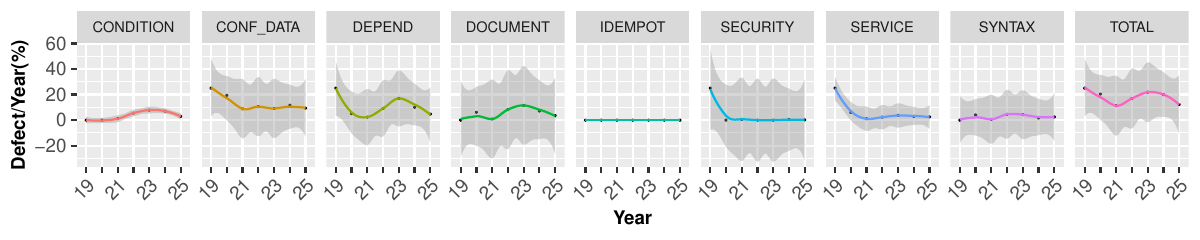}
  \label{fig:vtex-defect-year}

  \textbf{NUBANK}
  \includegraphics[width=\linewidth]{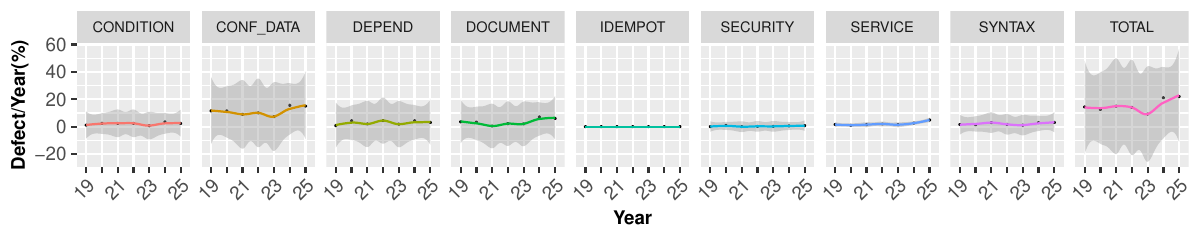}
  \label{fig:nubank-defect-year}

\end{figure*}

In Figure~\ref{fig:evolution-defect-year}, we present the Defect/Year metric across our datasets. The x-axis shows the years (last two digits), while y-axis displays the number of defects identified. Although the confidence interval (gray area) may dip below zero due to statistical estimation, the defects/year metric cannot be negative. Defects related to configuration data persist every year, as their defect-per-year values never reach zero—suggesting these issues are consistently reported across the entire lifespan of all three datasets.

We confirmed the presence of idempotency defects as an IaC-specific category, consistently appearing at low frequencies across all projects. Similarly, security defects were infrequent but persistent, in line with the findings of the original study. 

Additionally, in the proprietary datasets (VTEX and Nubank), internal developers reported robust quality controls that proactively mitigate defects. When asked about the defect/year metric, they noted that these controls likely contribute to the low incidence of idempotency and security-related issues.


\subsection{How do the frequency and distribution of PL-IaC defect types compare with those observed in the original study?}

\subsubsection{Method}
To answer RQ2.1, we conduct a straightforward comparison of the results for defect proportion and script proportion metrics between the two studies. We then discuss significant differences and similarities in defect distributions between our datasets and the original findings. 

\subsubsection{Answering RQ 2.1}

Among all defect categories, Configuration Data defects are the most frequent in all three datasets based on the defect and script proportion. While their frequency may be lower than in the original study, when comparing the script proportion metric, the dominance of this category remains consistent in both studies. Notably, within Configuration Data, the Network subcategory is also the most frequent, mirroring the original findings. Similar to the original study, defects related to Configuration Data persist every year, never reaching zero occurrences.

Despite differences in project types and technologies setup, \textbf{We observed highly similar distributions of defect proportions, with the original study reporting an average of 15.3, compared to 16.8 in our replication.} This consistency suggests that the original finding generalizes beyond Puppet and OSS ecosystems, aligning with the original authors' hypothesis that the frequency and distribution of defects transcends specific IaC tools. 

When correlating categories, our evolution on ACID tool demonstrated a higher number of correlated categories. We also detected rare cases of five-category overlaps (1.80\%, 0.63\%, 0.45\%), six-category overlaps in PIPr and Nubank only (0.43\%, 0.06\%), and even a small number of seven-category overlaps in PIPr (0.22\%). These results contrast with the original study, which reported at most three-category ECMs.

\section{Replication Package}
\label{sec:package}

We made a replication package available at GitHub\footnote{https://github.com/anonymous705/replication-package} and archived on Zenodo~\cite{zenodo}. This package includes the PIPr dataset, scripts, and intermediate results used in our study.
The repository includes a step-by-step guide for reproducing the findings presented in this paper. We have also provided configurations needed to replicate our experimental environment.

\section{Threats to Validity}
\label{sec:threat}

We built our Oracle to validate ACID through 22 independent raters. Despite replicating the number of raters from the original study and recruiting DevOps practitioners at the graduate and undergraduate levels, the heterogeneity in their experience levels with IaC and, specifically, PL-IaC, represents a potential threat. This limitation is mitigated through the implementation of a balanced incomplete block design, which ensures each is reviewed by at least two raters. Disagreements in the initial classification were resolved through arbitration by the first author, with the objective of enhancing the Oracle.

We acknowledge a potential threat to internal validity in our taxonomy construction due to the researchers' familiarity with the original study. However, this familiarity helped ensure faithful replication of the methodology, which is crucial for assessing framework generalizability. To mitigate bias, we implemented safeguards: independent initial coding, strict descriptive coding procedures, and validation by 22 independent raters with varying experience in IaC. The consistency between our findings and the original study further supports the validity of our approach.

A potential threat to validity is the tool's reliance on string patterns, which may cause false positives or negatives. Also, using commit messages to identify and categorize defects presents a potential threat due to variability in their quality and detail. Ambiguous or incomplete messages may hinder the ACID tool's ability to accurately determine defect categories. To address this, we validated the tool using an oracle and conducted sanity checks in both the qualitative and empirical studies. These procedures showed high precision and recall, reinforcing the reliability of our findings. 

\section{Conclusion}
\label{sec:conclusion}
In this study we replicated and validated the 'Gang of Eight' defect taxonomy proposed by Rahman et al.~\cite{go8} across PL-IaC tools (Terraform, Pulumi, AWS CDK) and diverse development contexts, including both extensive open-source (PIPr) and large-scale proprietary (VTEX, Nubank) projects. Our analysis confirmed the generalizability of the original eight defect categories (Cloud/Service, Configuration Data, Dependency, Documentation, Idempotency, Logic/Conditional, Security, and Syntax), demonstrating their relevance beyond the original study's focus on Puppet scripts.  From our analysis, we believe the Resource and Panic subcategories can provide additional detail to Service defects. However, in our view, this refinement does not represent a departure from the original taxonomy, but rather a proposed extension of it.

In terms of defect frequency and distribution, the close similarity of defect proportion between the original study (15.3) and our replication (16.8) supports the generalizability of the original findings beyond Puppet and open-source ecosystems. Configuration Data defects are consistently frequent across all datasets, while Idempotency and Security defects, although less common, persisted as notable concerns.

This consistency suggests that the fundamental challenges identified by the original work are not confined to specific tools or organizational settings but are inherent aspects of managing infrastructure through code. Our study provides empirical evidence supporting the hypothesis that these defect patterns transcend technological and organizational boundaries, highlighting core difficulties within the IaC paradigm itself.

As future work, we understand that it is important to get a deeper investigation into the root causes and specific fixing patterns for the highly prevalent Configuration Data defects within the PL-IaC paradigm. Understanding why these errors persist despite the use of general-purpose programming languages could lead to targeted prevention strategies and improved tooling. Furthermore, assessing the criticality or severity of these different defect types might be important to understanding their practical impact and enabling effective prioritization. 

The evolution of the ACID tool highlights the potential for automated defect detection, but also its current reliance on pattern matching. Future work could focus on developing different strategies, such as static analysis techniques or machine learning models for more accurate defect detection directly from the source code. Expanding the analysis to compare defect distributions and characteristics across different programming languages used within PL-IaC (e.g., TypeScript vs. Python vs. Go) could also yield valuable insights into language-specific challenges. Finally, exploring the feasibility of automated program repair techniques for common IaC defects presents another interesting research direction.

\section{Acknowledgment}
We thank the 22 participants who contributed to this study by manually classifying defects.

\bibliographystyle{plain}
\bibliography{bibs} 

\end{document}